\documentstyle[epsfig]{aipproc}
\font\small=cmr10
\font\smalli=cmmi10
\begin{document}
\title{Quantum Fluctuations of Light:\\
A Modern Perspective on\\ Wave/Particle Duality}

\author{H. J. Carmichael}
\address{Department of Physics, University of Oregon,
Eugene, Oregon 97403-1274}

\maketitle

\begin{abstract}
We review studies of the fluctuations of light made accessible by the
invention of the laser and the strong interactions realized in cavity QED
experiments. Photon antibunching, advocating the discrete (particles), is
contrasted with amplitude squeezing which speaks of the continuous (waves).
The tension between particles and waves is demonstrated by a recent
experiment which combines the measurement strategies used to observe
these nonclassical behaviors of light [Phys.~Rev.~Lett. {\bf 85},
3149 (2000)].
\end{abstract}

\section{Introduction}

Our meeting celebrates one hundred years of the quantum. It was in October of
1900 that Planck stated his law for the spectrum of blackbody radiation for the
first time publicly \cite{planck1}, and it was in December of that year that
he presented a derivation of that law in a paper to the German Physical Society
\cite{planck2} in which he states: ``We consider, however---this is the most
essential point of the whole calculation---$E$ to be composed of a very
definite number of equal parts and use thereto the constant of nature $h=6.55
\times 10^{-27}\mkern2mu\hbox{erg sec}$.'' Planck introduced his ``equal parts''
so that he might apply Boltzmann's statistical ideas to calculate an entropy.
The ideas required that he make a finite enumeration of states and hence the
discretization was necessitated by the statistical approach. The surprise for
physics is that fitting the data required the discreteness to be kept while the
more natural thing would be to take the size of the ``parts'' to zero at the
end of the calculation.

The story of Planck's discovery and what may or may not have been his
attitude to the physical significance of the persisting discreteness is one
to be told by others \cite{terhaar1,pais1,murdoch1,heilbron}. Looking backwards
with the knowledge of a physicist trained in the modern era, the essence of
the blackbody calculation is remarkably simple and provides a dramatic
illustration of the profound difference that can arise from summing things
discretely instead of continuously---i.e., making an integration. Mathematically,
the difference is almost trivial, but why the physical world prefers a sum over
an integral still escapes our understanding. Going, then, to the heart of the
matter, the solution to the blackbody problem may be developed from a
calculation of the average energy of a harmonic oscillator of frequency $\nu$
in thermal equilibrium at temperature $T$. Taking a continuous energy variable
$y=E/h\nu$ leads to the calculation 
\begin{equation}
\frac{\int_0^\infty ye^{-yx}dy}{\int_0^\infty e^{-yx}dy}=
-\frac{(1/x)^\prime}{1/x}=\frac1x,
\label{eqn:integral1}
\end{equation}
where we define
\begin{equation}
x\equiv\frac{h\nu}{kT},
\end{equation}
and $k$ is Boltzmann's constant ($\,{}^\prime$ denotes differentiation with respect
to $x$); the result is the one expected from the classical equipartition theorem. Taking
a discrete energy variable, $E_n=nh\nu$, gives
\begin{equation}
\frac{\sum_{n=0}^\infty ne^{-nx}}{\sum_{n=0}^\infty
e^{-nx}}=-\frac{[1/(1-e^{-x})]^\prime}{1/(1-e^{-x})}=\frac1{e^x-1}.
\label{eqn:sum1}
\end{equation}
There is agreement between the sum and the integral for $x\ll1$, when the average
energy is made up from very many of Planck's ``equal parts;'' this is the domain
of validity of the Rayleigh-Jeans formula. Outside this domain discreteness brings
about Planck's quantum correction.

The pure formality of the difference is striking. The blackbody problem hardly
demands that we take the energy quantum $h\nu$ too seriously, whether the
oscillator be considered to be a material oscillator or a mode of the radiation
field \cite{debye}. Indeed, summed over oscillators of all frequencies, the total
energy is in effect continuous still. Considering the radiation oscillators, it has,
in fact, only recently become possible to make a direct observation of discrete
single-mode energies. Remarkably, measurement are made at microwave frequencies
where the energy quantum is exceedingly small. The feat is accomplished using
the strong dipole interaction between an atom excited to a Rydberg state and a
mode of a superconducting microwave cavity cooled near absolute zero \cite{brune};
the system has become a paradigm for studies in cavity QED \cite{berman}. In this
talk I will discuss other results from the interesting field of cavity QED, but
for a physical system where the material and radiation oscillators have optical
frequencies.

So Eqs.~(\ref{eqn:integral1}) and (\ref{eqn:sum1}) contrast the discrete with the
continuous. Of course, concepts of both characters enter into classical physics:
Newton's mass point is discrete, the particle or atom is discrete, as are any
``things'' counted $\ldots$ on the other hand the evolution over time unfolds
continuously, the {\it location\/} of a particle lies in a continuum, Maxwell's
waves are continuous. The important point about classical thinking is that ideas
on the two sides remain apart from one another, even if they have sometimes
competed, as, for example, in the varied attempts to account for the nature of
light. Quantum physics on the other hand, as it developed in the three decades
after Planck's discovery, found a need for an uncomfortable fusion of the discrete
and the continuous. Arguments about particles {\it or\/} waves gave way to a
recognized need for particles {\it and\/} waves. Thus, throughout the period of the
old quantum theory, from Planck until Heisenberg \cite{heisenberg} and Schr\"odinger
\cite{schroedinger}, a genuine ``wave/particle duality'' steadily emerged. The full
history is complex \cite{stuewer,wheaton,murdock2} and I will mention only some of
the most often quoted highlights.

 Einstein, in a series of celebrated papers, lay down the
important markers on the particle side \cite{pais2}. Amongst other things, he
brought Planck's quantum into clear focus as a possible particle of light
\cite{einstein1}, argued that discreteness was essential to Planck's derivation
of the radiation law \cite{einstein2}, and incorporated the quantum and its
discreteness into a quantum dynamics which accounted for the exchange of energy
between radiation and matter oscillators in a manner consistent with that law
\cite{einstein3}. Adding to this, Bohr's work connecting Planck's ideas to
fundamental atomic structure must be seen to support an argument on the particle
side \cite{bohr1}. Yet Bohr, like most others, was opposed to Einstein's
tinkering with the conventional description of the free radiation field as a
continuous wave.

The case on the side of the wave was easily made on the basis of interference
phenomena. Nonetheless, over time it became clear that the particle idea could
not simply be dismissed and it was suggested that the clue to a union lay, not
in the nature of free radiation, but in the nature of the interaction of radiation
with matter. Planck was among those to expressed this view \cite{planck3}:
``I believe one should first try to move the whole difficulty of the quantum theory
to the domain of the interaction of matter with radiation.'' The suggestion was
followed up most seriously in a bold work authored by Bohr, Kramers, and Slater
(BKS) \cite{bohr2} and based on a proposal of Slater's \cite{slater1}. The key
element was not a part of Slater's original proposal, however, which had waves---a
``virtual radiation field''---guiding light particles \cite{slater2}. For BKS
there were no light particles. The virtual waves comprised the entire radiation
field, radiated {\it continuously\/} by virtual material oscillators. The response
of the matter to the continuous radiation obeyed a quite different rule though;
following Einstein's ideas on stimulated emission and absorption \cite{einstein3},
the wave amplitude was to determine probabilities for {\it discrete\/} transitions
(quantum jumps) between stationary states \cite{note1}. The aim was to retain both
the continuity of Maxwell waves and the discreteness of quantized  matter by confining
each to its own domain.

There was a price to be paid for preserving the apartness, however. The BKS scheme
was noncausal (stochastic) at a fundamental level and although energy and momentum
were conserved on average, they would not be conserved by individual quantum events.
Statistical energy conservation had been considered before; Einstein was one of
those who had toyed with the idea \cite{pais3}. BKS cast the idea in a concrete form
with predictions that would be tested within less than a year. Their proposal was
not entirely misguided; we meet with a ``virtual'' radiation field---though
mathematically more sophisticated---in modern field theory. The fatal weakness was
that their scheme did not causally connect the downward jump of an emitting atom
with the subsequent upward jump of a particular absorbing atom. Direct correlation
between quantum events was therefore excluded, yet correlation was just what the X-ray
experiments of Bothe and Geiger \cite{bothe} and Compton and Simon \cite{compton}
observed.

Quantum optics took up the theme of correlations between quantum events in the 1970s,
as lasers began to be used for investigating the properties of light. This talk
reviews a little of what has grown from those beginnings. Only a small piece of the
history is covered since the main story I want to tell is about a particular experiment
in cavity QED \cite{foster1}. The experiment uncovers the tensions raised by
wave/particle duality in a unique way, by detecting light as both particle and wave,
correlating the measured wave property (radiation field amplitude) with the particle
detection (photoelectric count). Thus, light is observed directly in both its character
roles, something that has not been achieved in a single experiment before.

We will work up to the new results gradually. We begin with an updated statement
of the BKS idea  (Sect.~\ref{sect:criterion}) which we use as a criterion to define
what we mean when speaking of the nonclassicality of light. I will then say a few
things about the cavity QED light source and what it is about cavity QED that makes its
fluctuations of special interest (Sect.~\ref{sect:source}). I first discuss the
fluctuations in their separate particle and wave aspects: photon antibunching, seen
if one correlates particle with particle (Sect.~\ref{sect:antibunching}), is contrasted
with quadrature squeezing which is seen if one correlates wave (amplitude) with wave
(Sect.~\ref{sect:squeezing}). Individually, photon antibunching and quadrature squeezing
each show light to be nonclassical by our criterion. Each  may be explained
however by modeling light as either purely particle or wave. Finally, I will describe
the wave-particle correlations measured by Foster {\it et al.\/} \cite{foster1}
(Sect.~\ref{sect:wave-particle}), where neither conception alone can explain what is observed.

\section{A  criterion for Nonclassicality}
\label{sect:criterion}
Although there are still a few contrary voices, the opinion amongst physicists generally
is that light---electromagnetic radiation at optical frequencies---must be quantized, with
the introduction of Einstein's light particle, in order to account for the full range
of observable optical phenomena. Einstein stated his view that something of the sort
might be the case in the introduction to his 1905 paper, where he writes \cite{einstein1}:
``One should, however, bear in mind that optical observations refer to time averages and
not to instantaneous values and notwithstanding the complete experimental verification
of the theory of diffraction, reflexion, refraction, dispersion, and so on, it is quite
conceivable that a theory of light involving the use of continuous functions in space
will lead to contradictions with experience, if it is applied to the phenomena of the
creation and conversion of light.''

Einstein identified specifically ``the phenomena of the creation and conversion of
light'' as the point where contradictions might be found. Considering modern quantum
optics experiments, it is indeed to the ``conversion'' or, more precisely, detection of
the light that we look to define what is, or is not, a failure of the classical wave theory.
Light is detected through the photoelectric effect where it is responsible, through some
process of conversion, for the appearance of countable events---i.e, the production of
photoelectric pulses. If the light is to be a continuous wave, it interfaces awkwardly
with the discreteness of the countable events. The BKS attempt at an interface is
nevertheless remarkably successful in accounting for the action of the light from most
sources on a detector. It is therefore commonly adopted as a criterion, or test, for
those phenomena that truly contradict classical ideas. It is adopted in the spirit of
Bohr's comment to Geiger after he had learnt of Geiger's new X-ray results~\cite{bohr3}:
``I was completely prepared [for the news] that our proposed point of view on the
independence of the quantum process in separated atoms should turn out to be
incorrect. The whole thing was more an expression of an attempt to achieve as great
as possible application of classical concepts, rather than a completed theory.''

\begin{figure}[b!]
\centerline{\epsfig{file=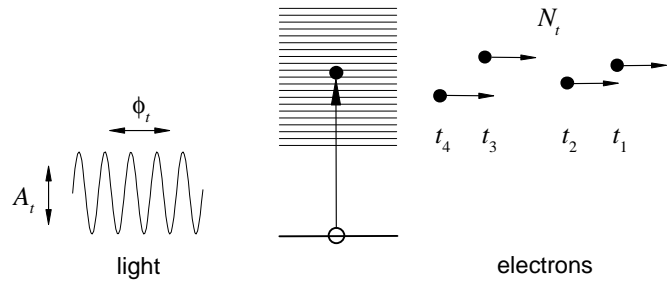,height=1.8in,width=3.6in}}
\caption{Semiclassical photoelectric detection of quasi-monochromatic light
couples a discrete stochastic process $N_t$ (photoelectron counting sequence)
to a continuous stochastic process $2A_t\cos(\omega_0t+\phi_t)$ (classical
electromagnetic field) through random detection events occuring, at time $t$,
at the rate $A_t^2$.}
\label{qfluct:fig1}
\end{figure}

Figure \ref{qfluct:fig1} illustrates the BKS interface as it is applied to the
photoelectric detection of light. On the left, the light is describe by a continuous
wave, specified at the position of the detector by an electric field, $2A_t\cos(\omega_0t
+\phi_t)$, whose amplitude $A_t$ and phase $\phi_t$ are generally fluctuating quantities
(random variables at each time~$t$). We consider the fluctuations to be slow compare to
the frequency $\omega_0$ of the carrier wave; thus, although the light has nonzero
bandwidth it is still quasi-monochromatic. On the other side of the figure, the sequence
of photoelectrons is discrete. Photoelectrons are produced at times $t_1,t_2,\ldots$, with
some number $N_t$ of them generated up to the time $t$. The difficulty is to interface
the continuity on the left with the discreteness on the right. This is done by allowing
the amplitude of the wave to determine the ``instantaneous'' rate at which the random
photoelectric detection events occur. With a suitable choice of units for $A_t$, the
probability per unit time for a bound electron to be released between $t$ and $t+dt$ is
given by the local time average of the light intensity, $A_t^2$.

The issue now is whether or not this model can account for what one observes with real
photoelectric detectors and light sources. Specifically, is it always possible to
choose a continuous stochastic process $(A_t,\phi_t)$ such that, in their statistical
properties (correlations), the observed photoelectric detection sequences which
constitute the experimental data can, in fact, be produced through the suggested
rule? The short answer, as we would expect, is that it is not always possible to do so.
On the other hand, for most light sources the BKS rule works just fine. It has actually
been quite an experimental challenge to produce light for which the rule fails.

\section{Light Sources and their Fluctuations}
\label{sect:source}
To start out we might ask how blackbody radiation fairs. By filtering a thermal
source, such as a spectroscopic lamp, it is possible to produce quasi-monochromatic
blackbody radiation. Consider then the fluctuations of such a light source. Let us
calculate the variance of the quasimode energy as Einstein first did in 1909
\cite{einstein4}. The continuous variable approach of Eq.~(\ref{eqn:integral1}) makes
the calculation appropriate to classical waves. Here we need the integration 
\begin{equation}
\overline{y^2}=\frac{\int_0^\infty y^2e^{-yx}dy}{\int_0^\infty e^{-yx}dy}=
\frac{(1/x)^{\prime\prime}}{1/x}=2\bar y^2,
\label{eqn:integral2}
\end{equation}
where $\bar y$ is the average energy, $1/x$; we obtain
\begin{equation}
\overline{\Delta y^2}\equiv\overline{y^2}-\bar y^2=\bar y^2.
\label{eqn:integral3}
\end{equation}
Alternatively, in the discrete variable approach of Eq.~(\ref{eqn:sum1}) we make
the summation 
\begin{equation}
\overline{n^2}=\frac{\sum_{n=0}^\infty n^2e^{-nx}}{\sum_{n=0}^\infty
e^{-nx}}=\frac{[1/(1-e^{-x})]^{\prime\prime}}{1/(1-e^{-x})}=2\bar n^2+\bar n,
\label{eqn:sum2}
\end{equation}
where $\bar n=1/(e^x-1)$ is the average energy. In this case we arrive at
\begin{equation}
\overline{\Delta n^2}\equiv\overline{n^2}-\bar n^2=\bar n^2+\bar n.
\label{eqn:sum3}
\end{equation}
Equation (\ref{eqn:sum3}) is the result obtained by Einstein and taken by him as
evidence that the theory of light would eventually evolve into ``a kind of fusion'' of
wave and particle ideas: light possess both a wave character, which gives the $\bar n^2$,
and a particle character, which gives the $\bar n$ \cite{einstein4,murdoch3,pais4}.

It would appear, then, that the detection of blackbody radiation would be incorrectly
described by the scheme of Fig.~\ref{qfluct:fig1}, since there the amplitude and energy
of the light wave is continuously distributed, which should lead to
Eq.~(\ref{eqn:integral3}), the incorrect result. This, however, is not the case at all;
thermal light fluctuations do not meet our criterion for nonclassicality. In fact BKS
made an attempt at the needed `fusion'. They did not eliminate particles to favor waves.
They attempted only to keep the particles and waves separate. The separation recovers the
two terms of Eq.~(\ref{eqn:sum3}) from two distinct (independent) levels of randomness.
To see this we must identify the integer $n$, not with the free radiation field, but with
the number of photoelectrons counted in a {\it measurement\/} of the field energy. The
first term, the wave-like term in Eq.~(\ref{eqn:sum3}), is then recovered from the
randomness of the field amplitude $A_t$, just as in Eq.~(\ref{eqn:integral3}), while the
second particle-like term is recovered from the addition randomness of the photoelectron
counting sequence introduced by the rule governing the production of photoelectrons.
Even if $A_t$ fluctuates not at all the photoelectron number will still fluctuate. It
will be Poisson distributed. The second term of Eq.~(\ref{eqn:sum3}) is recovered as
the variance of the Poisson distribution (which equals its mean).

Laser light is a good approximation to the ideal, coherent Maxwell wave which produces
only the Poisson fluctuations generated in the detection process. Of course, once one
has a laser, one can make a whole range of fluctuating light sources by imposing noisy
modulations of one sort or another. So long, however, as the fluctuations are imposed,
and thus independent of the randomness introduced in generating the photoelectrons,
nothing more regular than a Poisson photoelectron stream will be seen. Here, then, is
the Achilles heel of the BKS approach; it permits only super-Poissonian photoelectron
count fluctuations. Once again, the limitation involves a discounting of correlations
at the level of single quantum events. To illustrate, imagine a light source in which
the emitting atoms make their quantum jumps from higher to lower energy at perfectly
regular intervals. In the particle view, the source sends out a regular stream of
photons, which, supposing efficient detection, yields a regular, temporally correlated,
photoelectron stream. Such a photoelectron stream is impossible in the BKS view; its
observation would meet our criterion for nonclassicality.

Any experimental search for the disallowed correlations must begin with a method for
engineering light's fluctuations on the scale of Planck's energy quantum. What one can
do is begin with laser light and scatter it, through some material interaction, to produce
light that fluctuates in an intrinsically quantum mechanical way. Coherent scattering is
of no use, since it looks just like the laser light---neither is incoherent scattering
in which the fluctuations simply arise from noisy modulations. The fluctuations must be
caused by the ``quantum jumpiness'' in the matter;  the experiment must be sensitive
enough to see the effects of individual quantum events. This is rather a tall order,
since if we have in mind scattering from a sample of atoms, say, the effect, on the
fluctuations, of what any one atom is doing is generally very small. Happily, cavity
QED comes to our aid.  

\begin{figure}[t!]
\centerline{\hbox{\epsfig{file=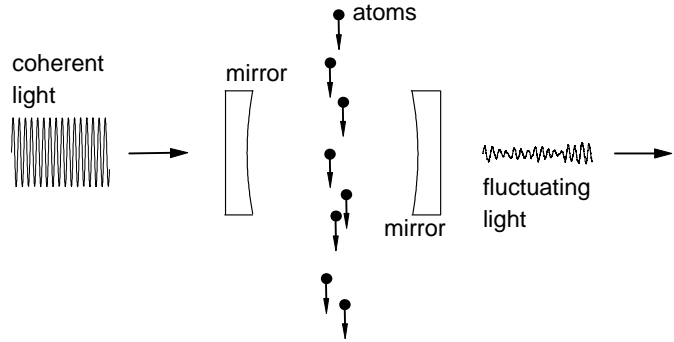,height=1.8in,width=3.6in}}}
\vskip0.1in
\caption{Schematic of the cavity QED light source. The input laser light
and Fabry-Perot cavity are both tuned to resonance with a dipole allowed
transition in the atoms.}
\label{qfluct:fig2}
\end{figure}

The light source used in the experiment I wish to discuss is illustrated in
Fig.~\ref{qfluct:fig2}. Basically, a beam of coherent light is passed through a
dilute atomic beam---at right angles to minimize the Doppler effect. The light is resonant
with a transition in the atoms, which make their ``jumps'' up and down while scattering
some of the light, and hence add fluctuations to the transmitted beam. The incoherent part
of this forwards scattering would be extremely small without the mirrors. They are
essential; they form a resonator which enhances the fluctuations. We might understand the
requirements for the resonator by observing that the goal (thinking of light particles) is
to redistribute the photons in the incoming beam. The interaction of the atoms with a
first photon must therefore change the probability for the transmission of the next. The
strength of any such collusion between pairs of photons is set by Einstein's induced
emission rate in the presence of a single photon. This rate must be
similar to the inverse residence time of a photon trapped between the mirrors. It follows
that the resonator must be small so that the energy density of one photon is large, and
the mirrors must be highly reflecting so that the residence time is long.

The experimental details go beyond the scope of this talk, but a few numbers might
be of interest \cite{rempe1,foster2}. Typical resonator lengths are 100-500$\mu$m with
50,000 bounces of a photon between the mirrors before it escapes. The transverse width
of the resonator mode is typically 30$\mu$m, which means the resonator confines a photon
within a volume of order $10^{-13}$m$^3$; the electric field of that photon is approximately
10$\mkern2mu{\rm V}{\rm cm}^{-1}$. The duration of a fluctuation written onto the light
beam may be estimated from the photon lifetime, $(L/c)N_{\rm bounce}\sim 50\mkern2mu$ns,
where $L$ is the resonator length and $N_{\rm bounce}$ is the number of mirror bounces.
This is a long time compared with the speed of modern photoelectric detectors  which makes
it possible to observe the fluctuations directly in the time domain. We should note, also,
that the fluctuations are extremely slow compared with the period of the carrier wave;
a typical fluctuation will last more than $10^7$ optical cycles.

It is not really necessary to understand what takes place inside the resonator. We are
interested in the results of measurements made on the output beam and whether or not
they can be reproduced by our detection model (Fig.~\ref{qfluct:fig1}) and {\it any\/}
fluctuating wave $A_t\cos(\omega_0 t+\phi_t)$. One feature of the data is particularly
noticeable though: an oscillation at a frequency of around $40\mkern2mu$MHz (see
Figs.~\ref{qfluct:fig4} and \ref{qfluct:fig9}), which suggests that the fluctuations
caused by the interaction with the atoms take the form of an amplitude modulation.
The modulation is a fundamental piece of phenomenology from the world of cavity QED,
referred to variously as a vacuum Rabi oscillation \cite{zhu,bernadot}, a normal-mode
oscillation \cite{raizen1}, or a cavity polariton oscillation \cite{weisbuch,stanley}.
The physics involved is rather simple. The electric field of the resonator mode excited
by the incident light obeys the equation of a harmonic oscillator, of frequency~$\omega_0$.
To a good approximation the electric polarization induced in the atoms by that field
is also described as a harmonic oscillator (Lorentz oscillator model), also with
frequency $\omega_0$. The two oscillators couple through the interaction between the
atoms and light; and coupled harmonic oscillators exchange energy coherently, back and
forth, so long as the period of the exchange---determined by the inverse of the
coupling strength---is shorter than the energy damping time. It is just this coherent
energy exchange that is seen in the fluctuations. The small mode volume of the resonators
used in cavity QED experiments ensures that the energy oscillation has a period shorter
than the damping time---although there are still some $10^7$ optical cycles during any
one period.

\section{Photon Antibunching: A Probe of Particle Fluctuations}
\label{sect:antibunching}
\begin{figure}[b!]
\centerline{\epsfig{file=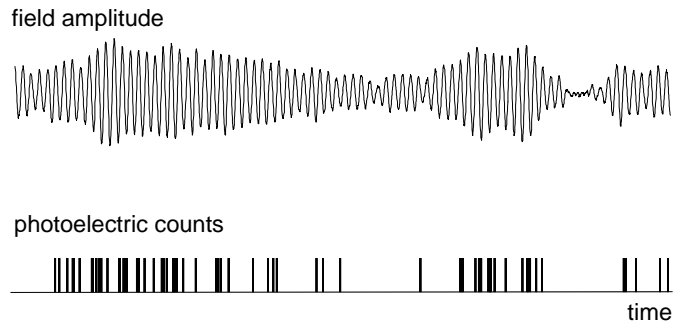,height=1.8in,width=3.6 in}}
\caption{Monte-Carlo simulation of a photoelectric count sequence produced,
with count rate $A_t^2$, by the shown fluctuating electromagnetic field
$A_t\cos(\omega_0t+\phi_t)$.}
\label{qfluct:fig3}
\end{figure}

Let us look first at a measurement that leads us towards the opinion that what is
transmitted by the resonator is a stream of light particles. In Fig.~\ref{qfluct:fig3},
we return, with more details, to our criterion for nonclassicality. Here, in a somewhat
arbitrary example, I have generated a realization of the photoelectric counts that
might be produced for a particular wave $A_t\cos(\omega_0t+\phi_t)$. It is of course
unreasonable to use a realistic carrier frequency, and therefore the frequency in
the picture is about a million times smaller---relative to the timescale of the
fluctuations---than it would be in reality.

We see from the figure that there are correlations between what the wave is doing and
the sequence of photoelectrons:  when the wave amplitude is large the photoelectric
counts come more quickly; when the wave amplitude is small there are gaps in the
count sequence. There are also correlations, over time, within each of the time
series. Thus, if at time $t$ the intensity $A_t^2$ is high, it is likely that
$A_{t+\tau}^2$ is also high for small positive and negative delays $\tau$. Equivalently,
if there is a photoelectron produced at time $t$ , there is a larger than average
chance that another is produced nearby at a delayed time $t+\tau$. Quantitative
statements about the correlations can be made by introducing the correlation function
\begin{equation}
g^{(2)}(\tau)\equiv\frac{\hbox{\small probability for a photoelectric detection at
times {\smalli t} and {\smalli t}+{\smalli\char28}}}{(\hbox{\small probability for a
photoelectric detection at time {\smalli t}})^{\mkern2mu 2}},
\end{equation}
where we will assume we are talking about stationary fluctuations, which simply means
that all probabilities and averages are independent of $t$; the correlation function
depends only on the time difference $\tau$.

\begin{figure}[t!]
\centerline{\epsfig{file=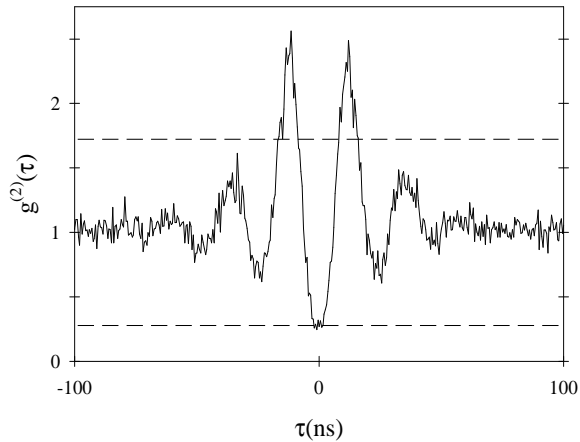,height=2.3in,width=3in}}
\caption{Violation of the inequalities imposed on the intensity correlation
function by the photoelectric detection scheme of Fig.~\ref{qfluct:fig1}.
To satisfy the inequalities the correlation function should show a maximum
greater than unity at zero delay; accepting the observed minimum, it must then
lie entirely between the dashed lines. Figure reproduced from Ref.~[34].}
\label{qfluct:fig4}
\end{figure}

According to the BKS detection model, photoelectrons are produced as random events at
rate $A_t^2$. The correlations in the photoelectron counting sequence are therefore
connected to the fluctuations of the wave through the joint
detection probability
\begin{equation}
\left\{\mkern-2mu\matrix{\hbox{\small probability for a photoelectric}\cr
\hbox{\small detection at times {\smalli t} and {\smalli t}+{\smalli\char28}}}
\mkern-2mu\right\}\propto A_t^2A_{t+\tau}^2;
\label{eqn:probability}
\end{equation}
there can be no stronger connection between events than can be expressed through
the correlations of the continuous variable $A_t$. It is rather easy to see
that this feature imposes constraints on the function $g^{(2)}(\tau)$. Specifically,
averaging over the random variables $A_t$ and $A_{t+\tau}$ one finds that the
inequalities
\begin{equation}
g^{(2)}(0)-1\geq0
\label{eqn:gtwo1}
\end{equation}
and
\begin{equation}
|g^{(2)}(\tau)-1|\leq|g^{(2)}(0)-1|
\label{eqn:gtwo2}
\end{equation}
must hold.  Inequality (\ref{eqn:gtwo1}), for example, restates a point we have already
noted; namely, that the BKS idea leads, unavoidably, to a photoelectron counting
sequence that is more irregular than a Poisson process. The inequality relies on
nothing more than the fact that the variance of $A_t^2$ must be positive. Needless-to-say,
to take over Einstein's expression \cite{einstein1}, inequalities (\ref{eqn:gtwo1}) and
(\ref{eqn:gtwo2}) ``lead to contradictions with experience.'' The data of
Fig.~\ref{qfluct:fig4}, as an illustration, violate both.

In order to obtain the result shown in the figure, the photoelectric counts must be
anticorrelated, in the sense that the appearance of a count at any particular time in
the photoelectron count sequence makes it less, rather than more likely, that another
will appear nearby. The phenomenon is called photon antibunching to contrast it with
the photon bunching---positive correlation---seen with a source of blackbody radiation
\cite{browntwiss}. The simplest example of antibunched light is provided by the
resonance fluorescence from a single atom \cite{carmichael1} and the first observations
of the phenomenon were made on the fluorescence from a dilute atomic beam \cite{kimble}.
More recently, beautiful measurements have been made on individual atoms, or more
precisely, electromagnetically trapped ions \cite{diedrich}. The data of
Fig.~\ref{qfluct:fig4} were taken for a cavity QED source like the one illustrated in
Fig.~\ref{qfluct:fig2} \cite{foster2}. Such a source produces a weak beam of
antibunched light \cite{rice,rempe2,mielke}.

Photon antibunching is nonclassical by our adopted criterion. It is incompatible with
the demarcation enforced by BKS between continuous light waves and discrete photoelectric
counts. It is quite compatible, on the other hand, with a particle constitution for
light. Indeed, there is nothing particularly peculiar, in principle, about a sequence
of photoelectric counts more regular than a Poisson process, and such a sequence could
be generated causally by a regular stream of light particles.

\section{Quadrature Squeezing: A Probe of\\ Wave Fluctuations}
\label{sect:squeezing}
The only difficulty with the stream of light particles is that, looked at in another
way, the same source of light does appear to be emitting a noisy wave. Whenever
interference is involved, a wave nature for light seems unavoidable. There are, of
course, numerous situations in which the interference of light is seen. We are all
familiar, for example, with Young's two-slit experiment. Considering wave aspects of
the fluctuations of light calls for an interference experiment that is just a little
bit more complex.

Balanced homodyne detection provides a method for directly measuring the amplitude
of a light wave. The method is carried over from the microwave domain and was proposed
in the 1980s \cite{yuen1} for detecting the fluctuations of what is known as quadrature
squeezed light \cite{stoler,lu,yuen2}. The light that produced Fig.~\ref{qfluct:fig4},
which is antibunched when photoelectron counts are considered, is quadrature squeezed
when its amplitude is measured. Like photon antibunching, quadrature squeezing
contradicts the BKS model of photoelectric detection;  according to our criterion it
is also nonclassical. It, however, leads us away from the stream of light particles
and towards the view that light is indeed a noisy wave; not, on the other hand, exactly
the wave BKS had in mind.

\begin{figure}[b!]
\centerline{\epsfig{file=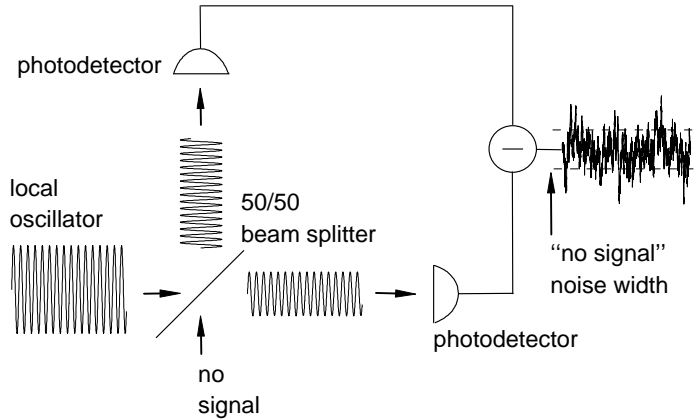,height=2.4in,width=3.6 in}}
\caption{In balanced homodyne detection, a coherent local oscillator field with
frequency $\omega_0$ matching that of the signal carrier wave is superposed
with the signal at a 50/50 beam splitter. The resulting output light is
detected with a pair of fast photodiodes and the two photocurrents subtracted to
zero the mean ``no signal'' current. An electronic shot noise remains, uncanceled;
it is necessarily present according to the scheme of Fig.~\ref{qfluct:fig1}
due to the randomness of the detection events that generate the photocurrents.
The ``no signal'' noise width measures the size of this noise and scales with the
amplitude of the local oscillator field and the square root of the detection
bandwidth.}
\label{qfluct:fig5}
\end{figure}

\begin{figure}[t!]
\hskip0.25in
\epsfig{file=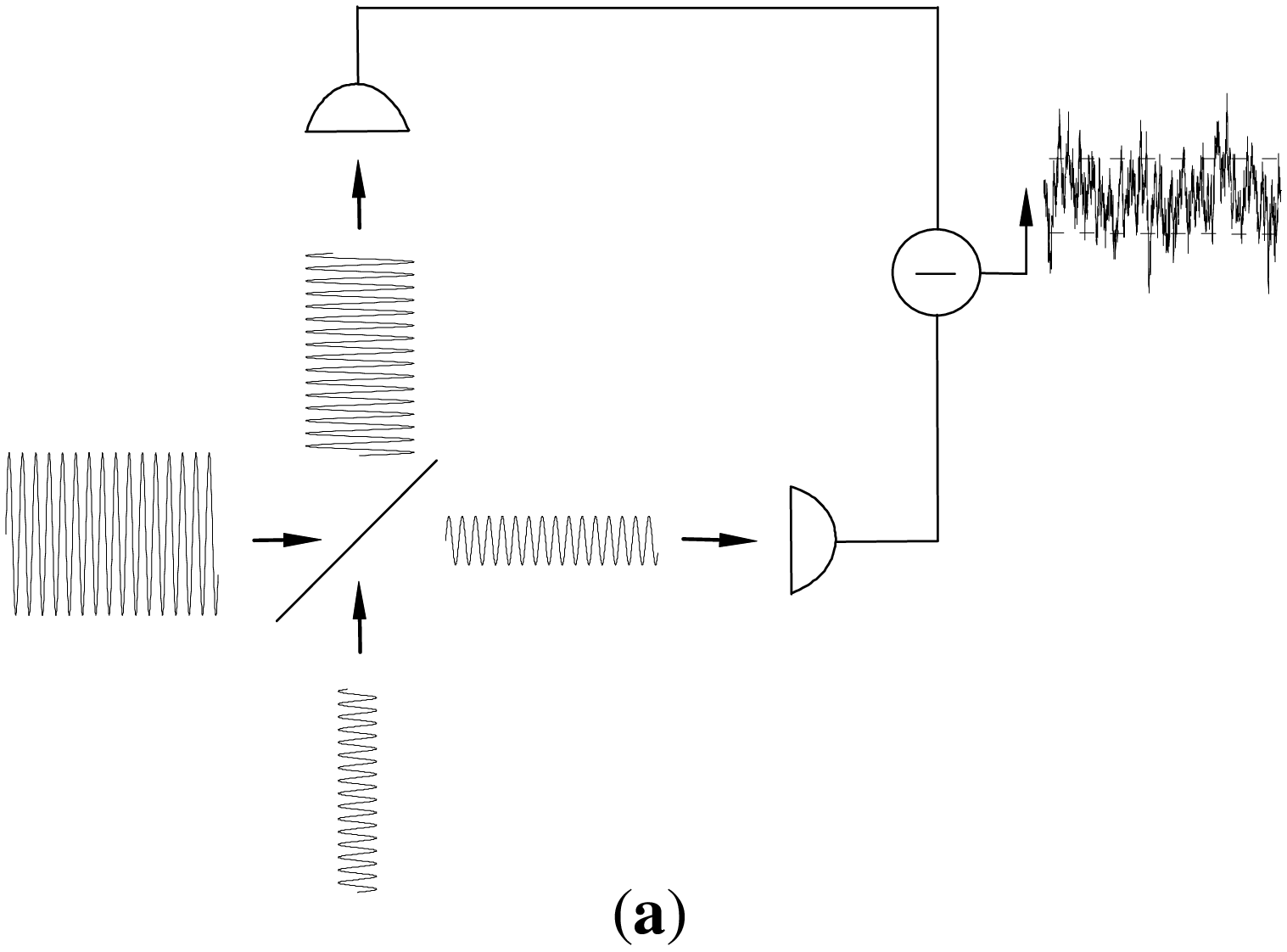,height=1.67in,width=2.5in}
\vskip-1.69in
\hskip3.0in
\epsfig{file=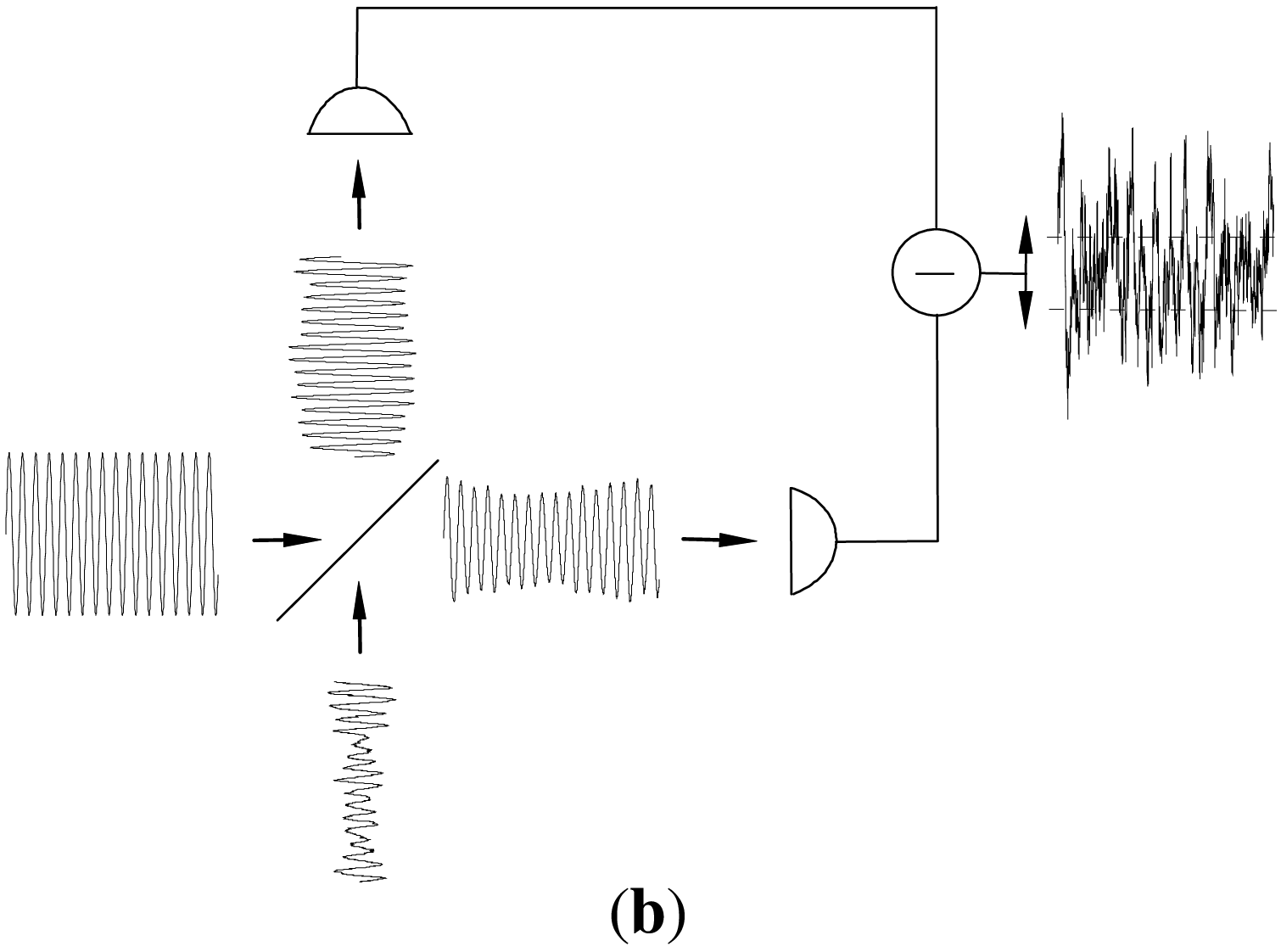,height=1.67in,width=2.5in}
\vskip0.4in
\caption{({\bf a}) A signal field of fixed amplitude and phase unbalances the
homodyne detector so that the mean difference current moves away from zero while
the noise width remains unchanged. ({\bf b}) A fluctuating signal unbalances the
detector in a noisy way, sweeping the difference current back and forth. This
introduces additional low-frequency noise which must {\it increase\/} the overall
noise width.}
\label{qfluct:fig6}
\end{figure}

I find it most helpful to understand quadrature squeezing in an operational way, so
I will proceed in this direction, and hopefully move ahead in a series of easy steps. The
basic idea in balanced homodyne detection is to interfere the signal wave, $A_t\cos(\omega
t+\phi_t)$, with a reference or local oscillator wave, $A_{LO}\cos(\omega t+\phi_{LO})$,
which ideally has a stable amplitude and phase. If the interference takes place at a 50/50
beam splitter as illustrated in Fig.~\ref{qfluct:fig5} (the signal wave is injected where
it says ``no signal''), then there are in fact two output fields,
\begin{mathletters}
\begin{eqnarray}
\hbox{\it field 1\/}&=&\frac1{\sqrt2}[A_{LO}\cos(\omega t+\phi_{LO})
+A_t\cos(\omega t+\phi_t)],\\
\hbox{\it field 2\/}&=&\frac1{\sqrt2}[A_{LO}\cos(\omega t+\phi_{LO})
-A_t\cos(\omega t+\phi_t)],
\end{eqnarray}
which respectively display constructive and destructive interference. These fields
are separately detected, and the rates at which photoelectrons are generated in the two
detectors, once again adopting the detection model of Fig.~\ref{qfluct:fig1}, are
\end{mathletters}
\begin{mathletters}
\begin{eqnarray}
\hbox{\it rate 1\/}&\approx&{\textstyle{1\over2}\displaystyle}[A_{LO}^2+A_{LO}A_t
\cos(\phi_{LO}-\phi_t)],
\label{eqn:ratea}\\
\noalign{\vskip2pt}
\hbox{\it rate 2\/}&\approx&{\textstyle{1\over2}\displaystyle}[A_{LO}^2-A_{LO}A_t
\cos(\phi_{LO}-\phi_t)],
\label{eqn:rateb}
\end{eqnarray}
where to obtain these expressions we square the fields, average over one carrier wave
period, and drop the term $A_t^2$ under the assumption $A_t\ll A_{LO}$. The average
photocurrents from the detectors are proportional to the rates (\ref{eqn:ratea}) and
(\ref{eqn:rateb}), and when the photocurrents are subtracted, the average difference
current provides a measurement of the amplitude $A_t$ (consider the case $\phi_{LO}=\phi_t$).
Thus, we have a device that measures the amplitude of a light wave and its operation
depends explicitly on the capacity of waves to interfere.

We now turn to the issue of fluctuations. Imagine first that there is no signal injected.
The two photocurrents are produced with equal photoelectron count rates, $\frac12A_{LO}^2$.
The average difference current is therfore zero. But according to the detection model
of Fig.~\ref{qfluct:fig1}, individual detection events occur randomly, and independently at
the two detectors; hence, the current fluctuates about zero. Since the counts are Poisson
distributed, there is a
\end{mathletters}
\begin{equation}
\hbox{\it ``no signal'' noise width\/}\propto\sqrt{{\textstyle{1\over2}\displaystyle}
A_{LO}^2+{\textstyle{1\over2}\displaystyle}A_{LO}^2}=A_{LO}.
\label{eqn:shot}
\end{equation}
This is an unavoidable background noise level and when a clean, noiseless signal is
injected, to unbalance the detector, as illustrated in Fig.~\ref{qfluct:fig6}(a), the
measurement of the signal amplitude is made against this background noise.

In the end, then, there is again a constraint, akin to inequalities (\ref{eqn:gtwo1}) and
(\ref{eqn:gtwo2}), imposed by the detection model. To see what it is, consider finally
the injection of a fluctuating signal. The signal adds a fluctuating offset, or unbalancing
of the detector, which sweeps the ``no signal'' noise band backwards and
forwards~[Fig.~\ref{qfluct:fig6}(b)] to produce a {\it larger\/} overall noise width;
for a fluctuating signal we must {\it add\/} the statistically independent
\begin{equation}
\hbox{\it signal noise width\/}\propto\sqrt{A_{LO}^2\overline{A_t^2}}=A_{LO}
\sqrt{\mkern2mu\overline{A_t^2}}
\end{equation}
to the ``no signal'' noise width, where $\overline{A_t^2}$ is the variance of the signal
fluctuations. Thus, according to the BKS detection model, measuring the fluctuations of the
light wave amplitude can never yield a noise width smaller than (\ref{eqn:shot}), the
width that in more conventional language is called the {\it shot noise level\/}.

In reality smaller noise can be seen, and is seen, for squeezed light. The first successful
experiment was performed in the mid-1980s \cite{slusher} and squeezing for a cavity
QED source like the one that produced the antibunched data of Fig.~\ref{qfluct:fig4} was
observed soon thereafter \cite{raizen2}. In general, the noise level is measured as a
function of frequency. It is therefore characterized fully by a spectrum of squeezing.
Figure \ref{qfluct:fig9}(c) shows an example for a cavity QED source. The squeezing occurs
around the frequency of the oscillation seen in Fig.~\ref{qfluct:fig4}.

Quadrature squeezing, like photon antibunching, reveals that a beam of light may exhibit
smaller fluctuations---more regularity---than is permitted by the random events that
make the interface between light waves and photoelectrons in Fig.~\ref{qfluct:fig1}.
In the case of photon antibunching, we may imagine that the regular photoelectrons
are seen because the light already, before interaction, possesses the discrete property
revealed in the photoelectron counting data---i.e., the light beam is itself a stream
of particles. With quadrature squeezing a similar tactic might be followed; the fluctuation
properties of the photocurrent might be transferred, ahead of any interaction with the
detector, to the beam of light. The one difficulty here, though, is that the injection of no
light also generates photocurrent noise, which is the situation depicted in
Fig.~\ref{qfluct:fig5}. The way around this obstacle is to say that a fluctuating wave
is present---call it the vacuum fluctuations---even in absolute darkness, and that it
is the interference of this ``noisy darkness'' with the local oscillator that is
responsible for the ``no signal'' noise width. A smaller noise level can then be seen
if one can deamplify the ``noisy darkness'' (vacuum fluctuations); the cavity QED system
of Fig.~\ref{qfluct:fig2} is a device that brings about deamplification.

I should stress that when one accounts for quadrature squeezing in this way, the
vacuum fluctuations need not be encumbered by any abstractions of modern
quantum field theory. The vacuum of radiation is literally filled with noisy waves,
precisely in the way proponents of stochastic electrodynamics assert it to be
\cite{marshall1,marshall2}.

\section{Wave-Particle Correlations}
\label{sect:wave-particle}
What we have seen so far amounts to a fairly traditional view on wave/particle duality,
although the players, photon antibunching and quadrature squeezing, are possibly
unfamiliar; photon antibunching sits comfortably on the particle side, while quadrature
squeezing, because of the role of interference, speaks for light as a wave. The recent
experiment by Foster et al.~\cite{foster1} brings the duality into focus in a more
perplexing way by putting both players into action at once. That is not to say that
it demonstrates a contradiction, of the sort that would be met~if, in a double-slit
experiment, one could record the choice, slit 1 or slit 2, for the path of every
particle, yet still observe an interference pattern on the screen. Nevertheless, data of
the discrete, particle-type, and continuous wave-type are taken simultaneously, so that
light is seen in the experiment to act as particle {\it and\/} wave. The experiment
underscores the subtlety involved in the coexistence of waves and particles under
Bohr's complementarity, the illusive contextuality of quantum mechanical explanations.
Specifically, the apparently satisfying explanations given for photon antibunching and
quadrature squeezing---passing whatever properties are seen in the data over to the
light---appear, in this wider context, to be something of a deception.

\begin{figure}[h!]
\centerline{\epsfig{file=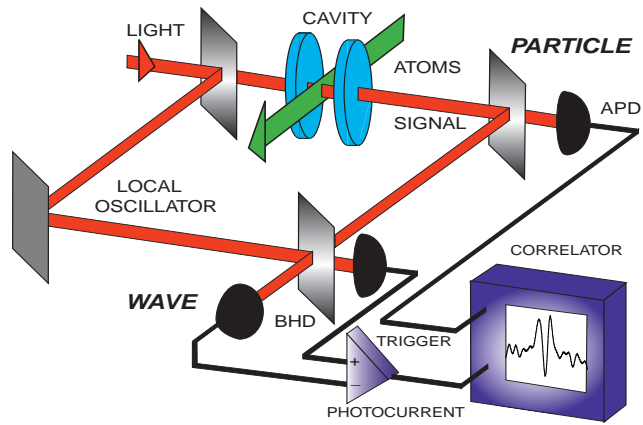,height=2.2in,width=3.3in}}
\caption{Experimental apparatus used to measure wave-particle correlations for the
cavity QED light source. Photoelectric detections at the avalanche photodiode
(APD) trigger the recording of the photocurrent from the balanced homodyne detector
(BHD). The correlator displays the cumulative average over many such records.
Figure reproduced from Ref.~[34].}
\vskip0.05in
\label{qfluct:fig7}
\end{figure}

The experimental apparatus is sketched in Fig.~\ref{qfluct:fig7}. At the top of the
figure there is a cavity QED system which acts as the source of fluctuating light.
The emitted light is divided between two detectors, one labeled PARTICLE which records
discrete photoelectric counts, and the other labeled WAVE is a balanced homodyne detector.
If all of the light were sent to just one detector, the apparatus could be used to measure
either photon antibunching or quadrature squeezing. In fact, the detectors are running
simultaneously. A count at the particle detector triggers the recording of the photocurrent
at the wave detector output---a little before and a little after the time of the
count---and many of these records are averaged to produce what appears on the oscilloscope.
What might we expect to see from this {\it conditional\/} measurement of the wave
amplitude? The experiment records the fluctuation of the amplitude of the wave that
accompanies the arrival of a photon at the particle detector. How will the wave and
particle properties be correlated?

\begin{figure}[b!]
\vskip-0.1in
\centerline{\epsfig{file=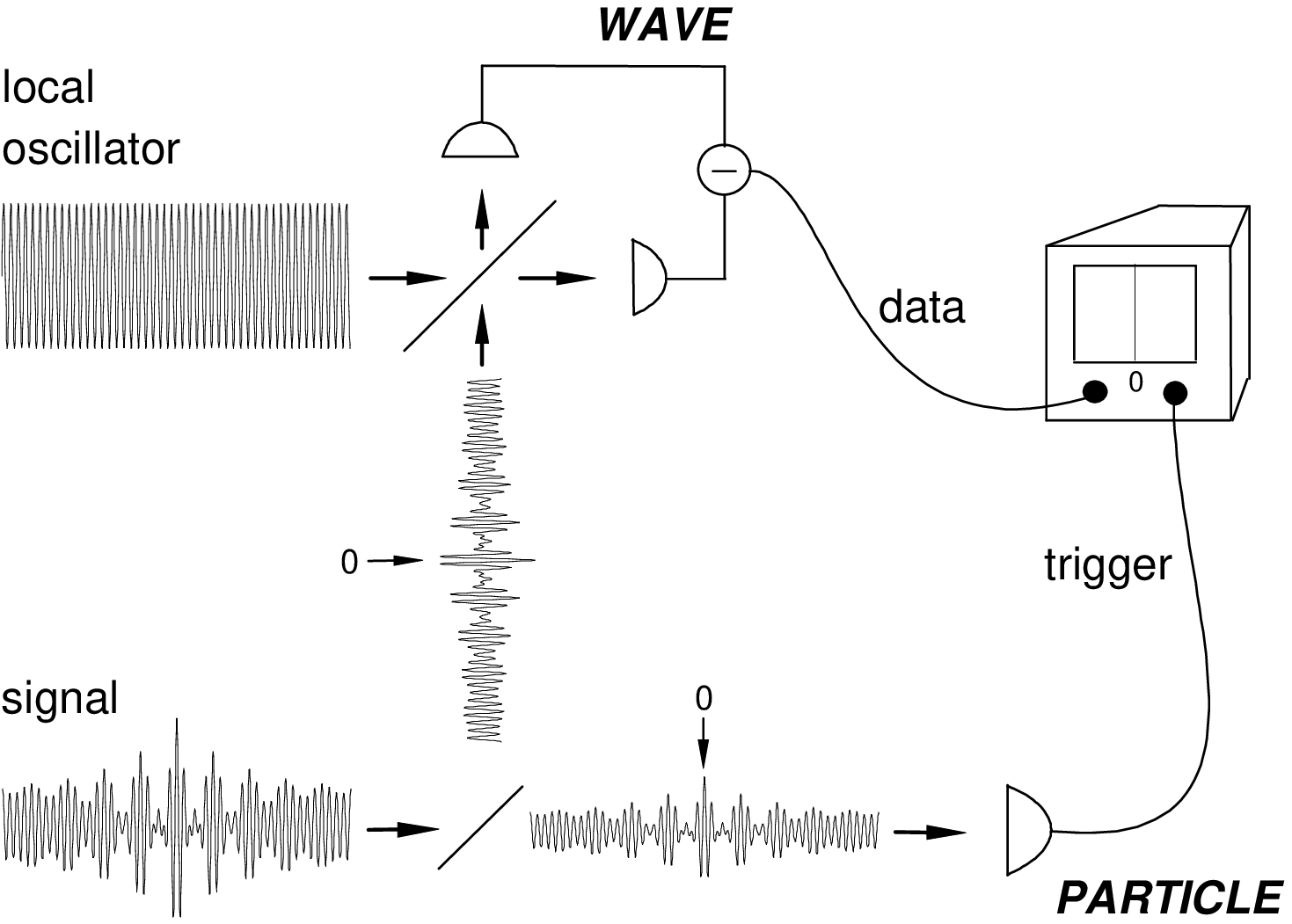,height=2.625in,width=3.5in}}
\vskip0.05in
\caption{Semiclassical analysis of the wave-particle correlator: The signal
fluctuation incoming from the lower left is divided at a beam splitter into two
parts, with one part sent to the particle detector and the other to the wave
detector. Each ``click'' of the particle detector fixes the time origin, $\tau=0$,
for a sampling of the wave detector output over the duration of the fluctuation;
the local oscillator phase is set to measure the amplitude of the wave
envelope. According to the photoelectric detection scheme of Fig.~\ref{qfluct:fig1},
the particle detector ``clicks'' most often when the intensity is {\it maximum\/}.
This should place the {\it maximum\/} of the measured wave amplitude at $\tau=0$.}
\label{qfluct:fig8}
\vskip0.2in
\centerline{\epsfig{file=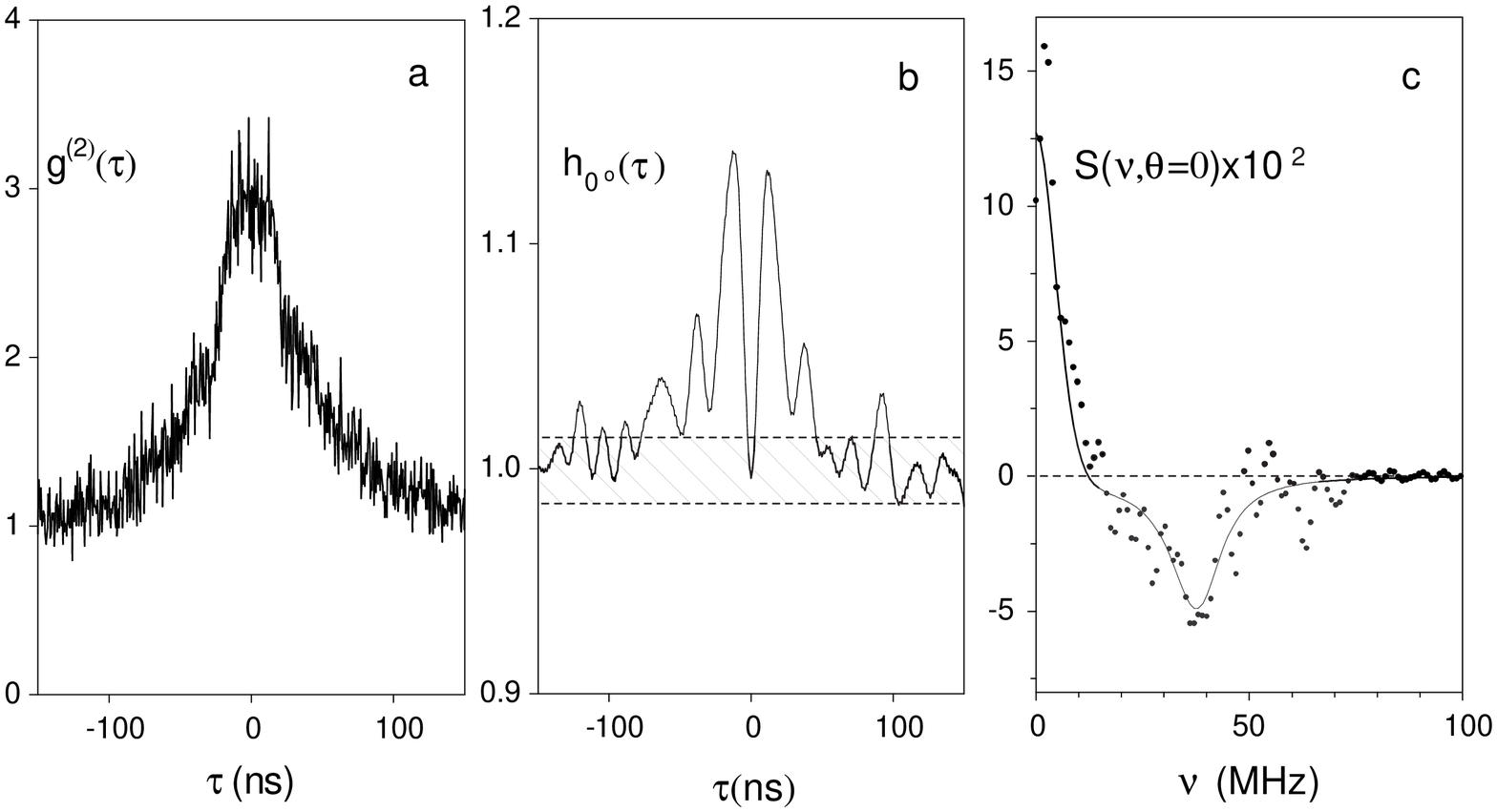,height=2in,width=4in}}
\vskip0.05in
\caption{Nonclassical wave-particle correlations for the cavity QED light source:
({\bf a}) the measured intensity correlation function is classically allowed,
({\bf b}) the corresponding wave-particle correlation function, which should lie
entirely within the shaded region according to the photoelectric detection scheme of
Fig.~\ref{qfluct:fig1}, ({\bf c}) the spectrum of squeezing obtained as the Fourier
transform of (b); for a classical field the spectrum would lie entirely above the
dashed line. Figure reproduced from Ref.~[34].}
\label{qfluct:fig9}
\end{figure}

In Fig.~\ref{qfluct:fig8} I attempt to show what would be expected on the basis
of the BKS detection model. A fluctuation from the light source is
injected into the correlator at the lower left; I give it the sort of amplitude
modulation evident in Fig.~\ref{qfluct:fig4}. The input wave is split, and passed on,
at half size, to the two detectors. Now the triggering sets the time origin for the
amplitude envelope function measured by the wave detector. The question, then, is, at
what point in time is the particle detector most likely to fire~$\ldots$ the answer:
when the fluctuation in the amplitude of the wave reaches its maximum. Strangely,
the reality is exactly the opposite, as is seen from the data shown in
Fig.~\ref{qfluct:fig9}(b). Figure \ref{qfluct:fig9}(a) shows the corresponding
particle-particle correlation function, $g^{(2)}(\tau)$, which in this case satisfies
the inequalities (\ref{eqn:gtwo1}) and (\ref{eqn:gtwo1}). In Fig.~\ref{qfluct:fig9}(c)
we see the spectrum of squeezing, which might have been
measured directly, but was in fact deduced from the correlation function plotted
in Fig.~\ref{qfluct:fig9}(b).

There are stronger signatures of nonclassicality to be observed than this conversion of
an expected maximum to a minimum. These may be stated quantitatively, as violations of
inequalities like those of Eqs.~(\ref{eqn:gtwo1}) and (\ref{eqn:gtwo2}) \cite{carmichael2}.
The most interesting says that the function plotted in Fig.~\ref{qfluct:fig9}(b) is
constrained under the BKS detection model by an absolute upper bound, $h_{\theta^\circ}
(\tau)\leq2$. The bound is not violated in the figure, but is predicted to be violated
in a more sensitive experiment by a factor of 10 or even 100. Considering the minimum
itself, though; how is it to be understood; and what does it have to say about the interplay
of waves and particles?

\begin{figure}[b!]
\hskip0.3in
\epsfig{file=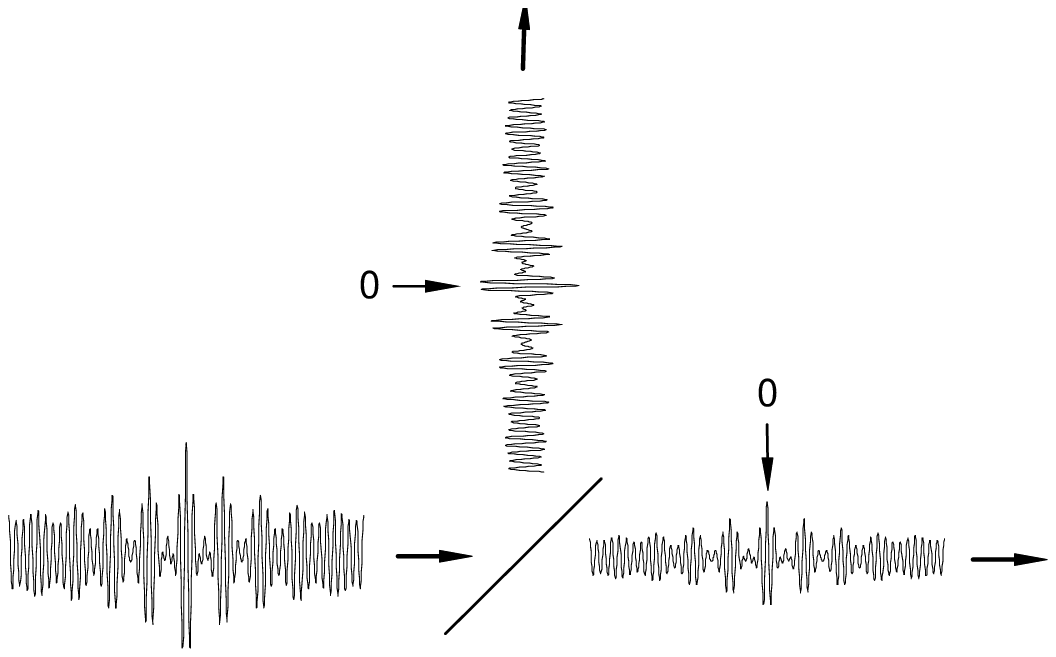,height=1.55in,width=2.5in}
\vskip-1.57in
\hskip3in
\epsfig{file=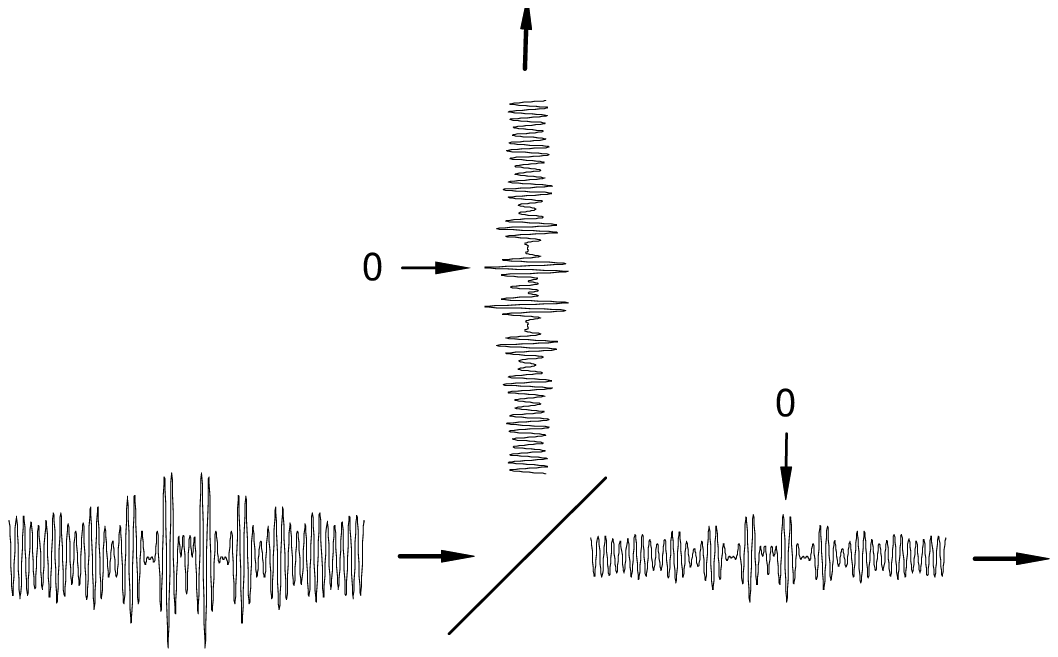,height=1.55in,width=2.5in}
\vskip.025in
\caption{Two possible signal fluctuations, centered, respectively, on a maximum
and a minimum in the amplitude of the wave envelope. In either case the particle
detector ``clicks'' most often on a maximum, which places the {\it maximum\/} of
the measured wave amplitude at $\tau=0$.}
\label{qfluct:fig10}
\end{figure}

\begin{figure}[t!]
\hskip0.3in
\epsfig{file=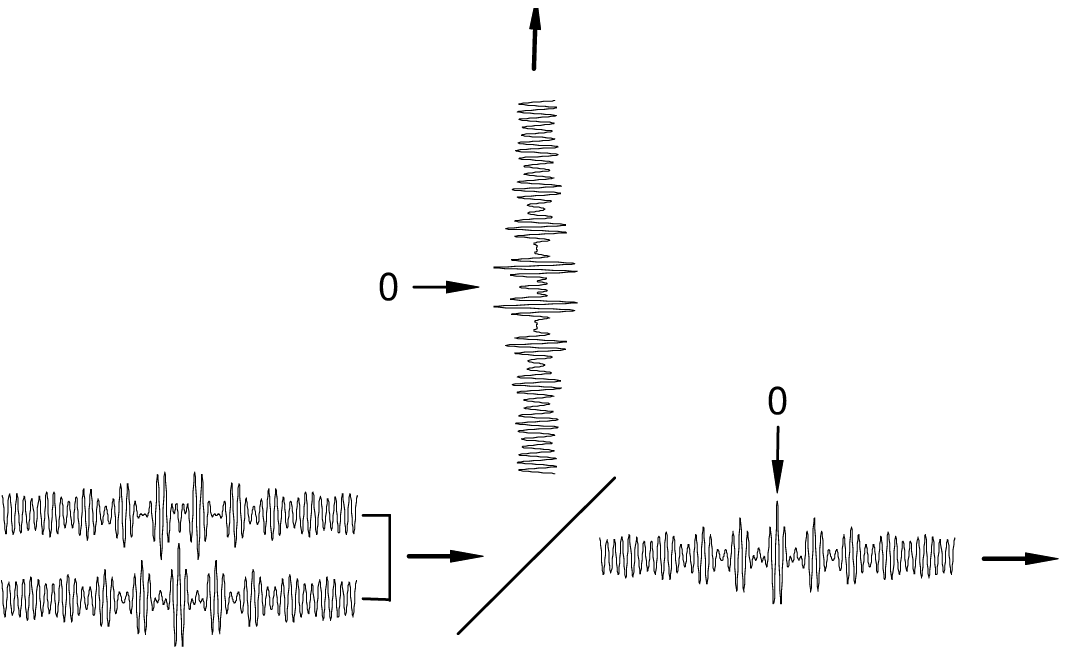,height=1.55in,width=2.5in}
\vskip-1.57in
\hskip3in
\epsfig{file=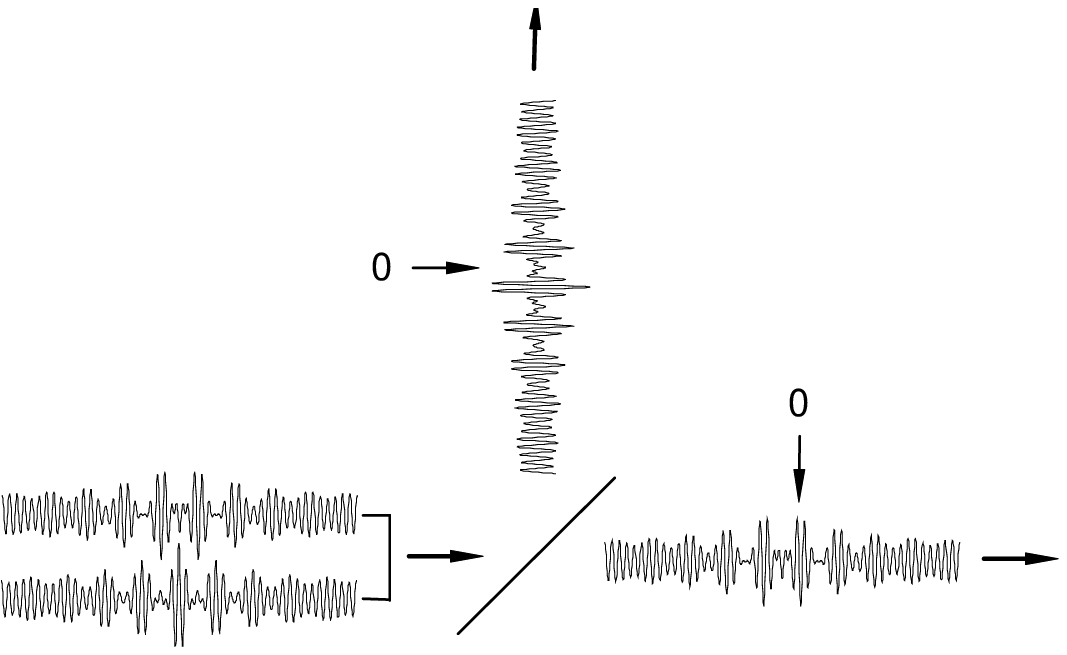,height=1.55in,width=2.5in}
\vskip.025in
\caption{Schematic illustration of how the anomalous wave-particle correlations
may be accounted for by combining wave and particle ideas. The alternatives in
Fig.~\ref{qfluct:fig10} are united as a single input fluctuation carried by a
two-particle wave with correlated maxima and minima. The beam splitter then
splits up the two-particle so as to preserve the wholeness of the individual waves.
For either of the splittings shown, the correlation between maxima and minima is
thus conveyed to the detectors so that the firing of the particle detector at the
intensity maximum places the measured wave amplitude {\it minimum\/} at $\tau=0$.}
\label{qfluct:fig11}
\end{figure}

Of course a calculation within the modern mathematical framework for treating
quantized fields predicts the minimum at $\tau=0$. Merely calculating gives little
physical insight though; for insight we turn to something more qualitative.
First, I should expand a bit on what is shown in Fig.~\ref{qfluct:fig8}. Over an
ensemble of triggered measurements, the phase of the modulated envelope function will
vary from shot to shot. Two extreme cases are shown in Fig.~\ref{qfluct:fig10}. In both,
according to the detection model of Fig.~\ref{qfluct:fig1}, triggering off a maximum
of the envelope places a maximum of the measured field amplitude at $\tau=0$---the
absolute locations of the maxima in the incoming fluctuations do not matter, only the
correlation between locations in the two waves emerging from the beam splitter. The
unexpected minimum of Fig.~\ref{qfluct:fig9}(b) may now be obtained, rather simply, by
viewing the cases shown, not as two possible outcomes realized on distinct occasions---one
{\it or\/} the other on any occasion---but as two possibilities that occur simultaneously,
and yet retain their distinctness. The words ``retain their distinctness'' are essential.
We are not to add together the waves shown bracketed in Fig.~\ref{qfluct:fig11} as one
would add classical waves. Each of these waves also has a discrete attribute, indicating
an individuality with respect to its counterpart, as a whole, distinct, ``one-particle
wave''---the two pieces being assembled, the bracketed object is a ``two-particle wave.''
In modern language we would call it a two-photon wavepacket.

To explain the data we now assert, that at the beam splitter, the discreteness, or
wholeness, comes into play, and one one-particle wave goes in either direction. We may
then continue with the idea that the particle detector is most likely to fire when the
intensity of the wave it sees is a maximum. With the now built~in anticorrelation of
modulation phases, whichever one-particle wave goes to the particle detector, the
amplitude recorded by the wave detector is at a minimum at the triggering time. The two
possibilities are shown in Fig.~\ref{qfluct:fig11}. There is of course a possibility
that both one-particle waves go to the same detector; occurences like this cannot,
however, upset the correlation recorded by the data.

Thus, if we are to account for the correlations observed with the apparatus of
Fig.~\ref{qfluct:fig7}, neither the particle stream that explained photon antibunching,
nor the noisy wave that explained quadrature squeezing will do. We need a composite
notion like the ``two-particle wave'' in order to embrace both pieces of the
correlation, both the discrete triggering event and the continuously measured amplitude.

\section{A Concluding Comment}
The main topics of my talk can be revisited in a short summary. We have seen how the
BKS idea embodied in the photoelectric detection model of Fig.~\ref{qfluct:fig1} is
unable to account for certain correlations exhibited by the fluctuations of light. In
these cases, an understanding of the correlations must relax the strict separation of
particle and wave concepts carried over by BKS from classical theory. Substituting
lasers for the blackbody sources studied by Planck, many experiments in recent years
have observed such nonclassical correlations. The experiment of Foster et al.
\cite{foster1} is notable, in particular, because if its simultaneous measurement, and
correlation, of the conflicting particle and wave aspects of light.

In a way, all of this serves only as an introduction to a second, more interesting,
story. Looking to the quantum mechanics that eventually emerged over the years after 
Planck, surely, now, we can give an unproblematic account of what light really is?
Unfortunately, in fact, we cannot, because we move here into new territory, where we have
to admit that although we have a formalism with which to calculate what we see, it is not
at all unbroblematic to put forward an ontology on which that formalism can rest. After BKS,
Bohr's thinking moved on to his ideas about complementarity \cite{bohr4,bohr5}. Einstein
never accepted these views, and on occasion dismissed them rather harshly \cite{einstein5}:
``The Heisenberg-Bohr soothing philosophy---or religion?---is so finely chiseled that it
provides a soft pillow for believers $\ldots$ This religion does dammed little for me.''
Thus, the second story, the enunciation of exactly where quantum mechanics has led us,
is an interesting one, but certainly also a difficult one to tell. All I can really do
at the conclusion of this talk, is indicate how the scheme of Fig.~\ref{qfluct:fig1}
is changed to give a unified, quantum mechanical description of the incoming light from
which the correct correlations can be extracted, whatever measurement is made.

\begin{figure}[b!]
\vskip-0.1in
\centerline{\epsfig{file=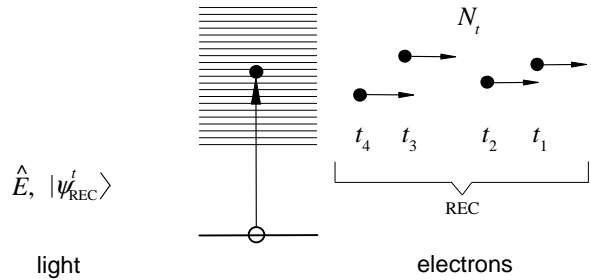,height=1.8in,width=3.6in}}
\vskip-0.05in
\caption{The quantum trajectory treatment of the photoelectric detection of light
couples the stochastic data record, $N_t$, to a stochastic state of the quantized
electromagnetic field, $|\psi_{\rm REC}^t\rangle$, through random detection
events occuring, at time $t$, at the rate $\langle\psi_{\rm REC}^t|\hat E^\dagger
\hat E|\psi_{\rm REC}^t\rangle$. The evolution of the quantum state becomes
stochastic because there is a state reduction $|\psi_{\rm REC}^t\rangle\to \hat E
|\psi_{\rm REC}^t\rangle$ (plus normalization) at the random times of the detection
events.}
\label{qfluct:fig12}
\end{figure}

The changed picture appears in Fig.~\ref{qfluct:fig12}
\cite{carmichael3,carmichael4,carmichael5}. I must point out two things. In change number
one, although, on the right, the photoelectrons are still conceived as a classical data
record, the light on the left is accounted for in abstract form. It is no longer a wave
of assigned quantitative value. It is represented by an {\it operator\/}, $\hat E$, which
is defined, not by a value but by the actions it might take; the value of the wave
emerges only when the operator acts---upon a second mathematical object, the state vector
$|\psi_{\rm REC}^t\rangle$. There is of course some mathematics that gives the explicit
forms of $\hat E$ and $|\psi_{\rm REC}^t\rangle$. For an appreciation of the scheme,
however, the mathematics is only a distraction.

Change number two, and an essential thing missing from the BKS proposal, is the label on
the state vector, REC. Through this label, the state of the incoming light is allowed to
depend on the history of the data record---the detection events that have already taken
place. At the time of each event, $\hat E$ acts on $|\psi_{\rm REC}^t\rangle$ to annihilate
a light particle, and in so doing updates the state of the incoming light to be consistent
with the record of photoelectric counts. In this way, correlations at the level of the
individual quantum events are taken into accounted. The communication through the label
REC is what, today, quantum physicists call back action, or in other words the reduction
of the state vector (or, less appealing, ``collapse of the wavefunction''), applied here to
the individual detection events. Without state reduction the Schr\"odinger equation
entangles the two sides of Fig.~\ref{qfluct:fig12}. It offers a nonlocal description in
terms of a global state vector. State reduction disentangles the state of the light from
the realized photoelectrons, and the correlations we have called nonclassical are indirect
evidence of this disentanglement.

Entanglement, nonlocality, state reduction, these are all words to remind us of the
problematic issue of ontology. Other speakers will talk about them more directly
\cite{shimony,omnes,wootters}. It is difficult to say what will come from the attention
these words are receiving one hundred years after Planck. Shall we come to understand
better, perhaps through a refinement of our faculties of perception following from the
amazing advances experimentation has made over these years. It might be appropriate to
take the final thoughts on the subject from Max Planck \cite{planck4}:

\noindent
$\ldots$``There is no doubt whatsoever that the stage at which theoretical physics has now
arrived is beyond the average human faculties, even beyond the faculties of the
great discoverers themselves. What, however, you must remember is that even if we
progressed rapidly in the development of our powers of perception we could not
finally unravel nature's mystery. We could see the operation of causation, perhaps,
in the finer activities of the atoms, just as on the old basis of the causal
formulation in classical mechanics we could perceive and make material images of
all that was observed as occurring in nature.

Where the discrepancy comes in to-day is not between nature and the principle of
causality, but rather between the picture which we have made of nature and the
realities in nature itself.''


\begin{references}
\bibitem{pais2} A complete list of Einstein's contributions with extensive
commentary is given by A.~Pais \cite{pais1}.
\end{references}
\end{document}